\documentstyle[amssymb,preprint,prb,aps,graphics,lscape]{revtex}

\tightenlines

\begin{document}
\title{Magnetic properties of the frustrated AFM spinel ZnCr$_2$O$_4$ and the
spin-glass Zn$_{1-x}$Cd$_x$Cr$_2$O$_4$ ($x=0.05,0.10$)}
\author{H. Martinho, N.O. Moreno, J.A. Sanjurjo, and C. Rettori}
\address{Instituto de F\'{i}sica\ ''Gleb Wataghin'', UNICAMP, 13083-970, Campinas-SP,%
\\
Brazil.}
\author{A.J. Garcia-Adeva and D. L. Huber}
\address{University of Wiconsin-Madison, Madison, WI, 53706, U.S.A}
\author{S. B. Oseroff}
\address{San Diego State University, San Diego, CA 92182, U.S.A.}
\author{W. Ratcliff II and \ S.-W Cheong}
\address{Lucent Technology, Bell Laboratories, Murray Hill, N.J. 07974, U.S.A.}
\author{P.G. Pagliuso and J. L. Sarrao}
\address{Los Alamos National Laboratory, Los Alamos, New Mexico 87545, U.S.A.}
\author{G. B. Martins}
\address{National High Magnetic Field Laboratory, Florida State University,\\
Tallahassee, FL 32306, U.S.A.}
\maketitle

\begin{abstract}
The $T$-dependence ($2$- $400$ K) of the electron paramagnetic resonance
(EPR), magnetic susceptibility, $\chi (T)$, and specific heat, $C_{v}(T)$,
of the $normal$ antiferromagnetic (AFM) spinel ZnCr$_{2}$O$_{4}$ and the
spin-glass (SG) Zn$_{1-x}$Cd$_{x}$Cr$_{2}$O$_{4}$ ($x=0.05,0.10$) is
reported. These systems behave as a strongly frustrated AFM and SG with $%
T_{N}$ $\ \approx T_{G}\approx 12$ K and $-400$ K $\gtrsim \Theta
_{CW}\gtrsim -500$ K. At high-$T$ the EPR intensity follows the $\chi (T)$
and the $g$-value is $T$-independent. The linewidth broadens as the
temperature is lowered, suggesting the existence of short range AFM
correlations in the paramagnetic phase. For ZnCr$_{2}$O$_{4}$ the EPR
intensity and $\chi (T)$ decreases below $90$ K and $50$ K, respectively.
These results are discussed in terms of nearest-neighbor Cr$^{3+}$ (S $=3/2$%
) spin-coupled pairs with an exchange coupling of $\left| J/k\right| \approx 
$ $50$ K. The appearance of small resonance modes for $T\lesssim 17$ K, the
observation of a sharp drop in $\chi (T)$ and a strong peak in $C_{v}(T)$ at 
$T_{N}=12$ K confirms, as previously reported, the existence of long range
AFM correlations in the low-$T$ phase. A comparison with recent neutron
diffraction experiments that found a near dispersionless excitation at $4.5$
meV for $T\lesssim T_{N}$ and a continuous gapless spectrum for $T\gtrsim
T_{N}$, is also given.
\end{abstract}


\section{\bf INTRODUCTION }

Frustration in the antiferromagnetic (AFM) ordering of fcc and spinel
lattices was recognized long ago by Anderson in his analysis of the
highly degenerate magnetic ground state of these structures.\cite{Anderson}
This so-called $geometrical$ frustration can prevent the system from undergoing 
spin-glass (SG) or AFM ordering down to temperatures much lower than the
Curie-Weiss temperature, $T_{G}$,$T_{N}<<\left| \Theta _{CW}\right| $. It
has also been shown, theoretically \cite{De Seze} and experimentally, \cite
{Fiorani} that the ground state degeneracy can be removed by atomic disorder
leading to a SG type of ordering. The AFM $normal$ spinel ZnCr$_{2}$O$_{4}$
structure, in which the octahedral Cr sites form corner-sharing tetrahedra, \cite{Barth} as well as the pyrochlore \cite{Moessner}$^{,}$\cite
{Canals} and kagom\'{e}\cite{Ramirez} structures, are excellent systems to
study the \emph{geometric} frustration phenomenon. ZnCr$_{2}$O$_{4}$ has a
very high Curie-Weiss temperature, $\Theta _{CW}\approx -400$ K, and a first
order AFM transition at $T_{N}\approx $ $12$ K \cite{Plumier} accompanied by
a slight tetragonal crystal distortion ($\Delta a/a\approx 10^{-3}$). \cite
{Hartmann} Besides, recent interesting low-$T$ neutron diffraction
experiments showed, in the ordered phase ($T\lesssim T_{N}$), the existence
of a near dispersionless excitation at $4.5$ meV, and for $T\gtrsim T_{N}$,
a continuous magnetic gapless density of states.\cite{S.-H. Lee}

In this work we report on the $T$-dependence ($2$ - $400$ K) of the electron
paramagnetic resonance (EPR), magnetic susceptibility, $\chi (T)$, and
specific heat, $C_{v}(T)$, in a single crystal of the $normal$ spinel ZnCr$%
_{2}$O$_{4}$ of cubic structure ($Fd\overline{3}m$, O$_{h}^{7}$) and in the
Cd doped polycrystalline Zn$_{1-x}$Cd$_{x}$Cr$_{2}$O$_{4}$ ($x=0.05,0.10$)
compounds. A polycrystalline isomorphous compound of ZnGa$_{2}$O$_{4}$ was
also used as a reference compound for the specific heat measurements.

\section{\bf EXPERIMENTAL DETAILS}

Single crystals of ZnCr$_{2}$O$_{4}$ of typical size of $\sim $2x2x2 mm$^{3}$
were obtained by the method of solid state reaction between stoichiometric
amounts of Cr$_{2}$O$_{3}$ and ZnO in air.\cite{Cheong} The crystals show
natural growing (001), (111), and (011) faces that were checked by the usual
Laue method. Polycrystalline Cd doped Zn$_{1-x}$Cd$_{x}$Cr$_{2}$O$_{4}$ ($%
x=0.05,0.10$) and ZnGa$_{2}$O$_{4}$ samples were prepared by the same
method. The EPR experiments were carried out in a conventional ELEXSYS
Bruker X-Band EPR spectrometers using a TE$_{102}$ room temperature cavity.
The sample temperature was varied by a temperature controller using helium
and nitrogen gas flux systems. This set up assures one that the spectrometer
sensitivity remains about the same over a wide range of $T$. Magnetization
measurements have been taken in a Quantum Design $dc$ SQUID MPMS-5T
magnetometer. The specific heat was measured using the heat pulse method in
a Quantum Design calorimeter using the QD-PPMS-9T measurement system.

\section{\bf EXPERIMENTAL RESULTS}

For the ZnCr$_{2}$O$_{4}$ single crystal, as the temperature is lowered from
room-$T$, the EPR line broadens and its intensity goes through a
maximum at about $90$ K with no measurable resonance shift. Figure 1 shows
the $T$-evolution of the EPR spectra between $4$ K and $45$ K. For $%
T\lesssim 17$ K the resonance distorts and small resonances modes emerge at
low-$T$. These modes do not depend on whether the EPR spectra are taken
under field cooling (FC) or zero field cooling (ZFC) conditions. But, they
depend on the size and shape of the sample and show a slight orientation
dependence (see inset). For $T\gtrsim 20$ K the EPR spectra show a single
isotropic resonance. For the Cd doped samples the resonance also broadens,
but the intensity increases down to $T\approx 12$ K (see below) where,
again, the resonance distorts and small resonance modes are seen. These
modes also show the same spectra under FC and ZFC conditions.

Figure 2 shows the $T$-dependence of the linewidth, $\Delta H_{1/2}$, and $g$%
-value between $18$ K and $400$ K for the crystal of Figure 1. The linewidth
broadens at low-$T$, the $g$-value is $T$-independent and its value, $g$ $=$ 
$1.978(5)$, corresponds to that of Cr$^{3+}$ ($3d^3$, S $=3/2$) ions $g$%
-value in a cubic site. \cite{Baran}$^{-}$\cite{Boom} Both, the $g$-value
and linewidth are isotropic for $T\gtrsim $ $20$ K. For the Cd doped samples
similar resonance line broadening and $g$-values are obtained (not shown).

Figure 3 presents the $dc$ magnetic susceptibility, $\chi (T)$, corrected by
the host diamagnetism in the range between $2$ K and $400$ K for the same
crystal of Figure 1. FC and ZFC measurements at $H$ $=2$ kOe and $10$ kOe
gave no difference for the susceptibility data. At low field $\chi (T)$
shows the typical 3D AFM ordering with $\chi (T\rightarrow 0)\approx $ $%
(2/3)\chi _{max.}(T\approx $ $45$ K). The inset shows the sharp drop of the
susceptibility at $T$ $=12(1)$ K. This temperature defines the N\'{e}el
temperature, $T_N$, for the 3D long range AFM ordering in ZnCr$_2$O$_4$. The
inset shows that, for $T\lesssim T_N$, the susceptibility is field
dependent, $\chi (T,H)$. This has been attributed to domain wall movement
in the AFM ordered state. \cite{D. Fiorani}

Figure 4 compares the magnetic susceptibility of the ZnCr$_{2}$O$_{4}$
single crystal of Figure 3 with those of the polycrystalline Zn$_{1-x}$Cd$%
_{x}$Cr$_{2}$O$_{4}$ ($x=0.05,0.10$) samples. For $T\gtrsim $ $100$ K the
data for the three compounds can be fitted to the usual Curie-Weiss law. The
linear fit yields to an effective number of Bohr magnetons $\mu _{eff}$ $%
=3.95(10)$ $\mu _{B}$, as expected for Cr$^{3+}$ ($g$ $=1.978$, S $=3/2$),
and a Curie-Weiss temperature, $\Theta _{CW}$ $=-390(20)$ K for ZnCr$_{2}$O$%
_{4}$. In a molecular field approximation $\Theta _{CW}$ $=$ S(S+1) $%
Jz/3k_{B}$. For S $=3/2$, $z$ $=6$ (nearest-neighbors), and $\Theta _{CW}$ $%
=-390$ K we obtain $J/k\approx $ $-50$ K. The Curie-Weiss parameters for the
Zn$_{1-x}$Cd$_{x}$Cr$_{2}$O$_{4}$ ($x=0.05,0.10$) samples are given in Table
I. For $T\lesssim $ $100$ K Figure 4 shows, however, that there is a
significant difference between the pure and Cd doped compounds. Low (high)
field ZFC-FC measurements show, in the Cd doped samples, the typical SG
irreversibility (reversibility) for $T\lesssim T_{G}\approx 12$ K (see inset
of Figure 4). These results and the large values found for $\mid \Theta
_{CW} $ $\mid $ (see Table I), clearly indicate that the Cd doped samples
develop a highly frustrated SG-type behavior with $T_{G}\approx 12$ K.

Figure 5 presents the $T$-dependence of the resonance intensity, $I(T)$, for
the ZnCr$_{2}$O$_{4}$ single crystal and the polycrystalline Zn$_{1-x}$Cd$%
_{x}$Cr$_{2}$O$_{4}$ ($x=0.05,0.10$) samples. Using an EPR standard, we
found that the intensity of the resonance at room-$T$, $I(300$ K),
corresponds to the total amount of Cr$^{3+}$ ions present in the samples.
Similar to the susceptibility data shown in Figure 4, here also we observe
two $T$-regimes, above and below $T\approx $ $100$ K. For $T\lesssim $ $100$
K $I(T)$ shows significant difference between the pure and Cd doped
compounds. For the Cd doped samples we found that $I(T)$ and $\chi (T)$
correlate well above $T_{G}\approx 12$ K (not shown); however, for the pure
sample this correlation is only observed for $T\gtrsim $ $100$ K (see inset
in Figure 5).

Figure 6 presents the $T$-dependence of the specific heat, $C_{v}(T)$, for
the ZnCr$_{2}$O$_{4}$ single crystal, the polycrystalline Zn$_{1-x}$Cd$_{x}$%
Cr$_{2}$O$_{4}$ ($x=0.05,0.10$) samples, and the reference compound ZnGa$%
_{2} $O$_{4}$. The inset show the strong effect that the Cd impurities have
on the AFM transition of the pure compound ZnCr$_{2}$O$_{4}$. The large
reduction in the peak of the $C_{v}(T)$ confirms the assignment of SG
character for the transition observed at $\approx 12$ K in the
susceptibility data for the Cd doped samples. The transition temperatures, $%
T_{G}$ and $T_{N}$, are in fairly good agreement with those extracted from
the susceptibility data (see inset in Figure 4). Fields up to $9$ T, within
the data resolution, did not affect $C_{v}(T)$ and the AFM and SG
transitions temperatures.

\section{\bf ANALYSIS AND DISCUSSION}

The above EPR and magnetic susceptibility results show that the cubic $%
normal $ spinel ZnCr$_{2}$O$_{4}$ and Zn$_{1-x}$Cd$_{x}$Cr$_{2}$O$_{4}$ ($%
x=0.05,0.10$) compounds present interesting magnetic behavior between $2$
K and $400$ K. A high-$T$ paramagnetic phase (HTPP), $T\gtrsim $ $100$ K for
ZnCr$_{2}$O$_{4}$ and $T\gtrsim $ $12$ K for the Cd doped samples, and a low-%
$T$ ordered phase (LTOP), $T\lesssim $ $12$ K, AFM for the pure and SG for
the doped compounds. For ZnCr$_{2}$O$_{4}$, a transition between these two
regimes is observed in the interval between $12$ K and $100$ K. Our high-$T$
EPR results are in general agreement with those already reported for
polycrystalline samples. \cite{Baran}$^{,}$\cite{Forni} However, the low-$T$
EPR data for our ZnCr$_{2}$O$_{4}$ single crystal are quite different from
those reported in Ref. 12. As the temperature decreases in the HTPP, the Cr$%
^{3+}$ magnetic moments experience short range AFM correlations. The
evidence for it is that, for $T\gtrsim $ $2T_{N}$, the EPR resonance shows
no $g$-shift and a $T$-dependence of the line broadening expected for a
short range magnetic interaction in AFM materials above the N\'{e}el
temperature $T_{N}$: \cite{Huber}

\begin{equation}
\Delta H_{1/2}(T)-\Delta H_{1/2}(\infty )=\frac{R}{[T-T_{N}]^{x}}  \label{1}
\end{equation}
The solid line in Fig. 2 shows the fitting of the data to Eq. 1. The fitting
parameters are: $x=1.12(1)$, $T_{N}$ $=12(1)$ K, $\Delta H_{1/2}(\infty )$ $%
=250(10)$ Oe, and $R$ $=110(20)$ Oe K$^{x}$. We should mention that, for $%
T\gtrsim T_{N}$, recent neutron diffraction measurements found a continuous
gapless spectrum that was attributed to quantum critical fluctuations of
small short range AFM correlated domains.\cite{S.-H. Lee} In the HTPP of ZnCr%
$_{2}$O$_{4}$, and for $T\gtrsim $ $100$ K, $\chi (T)$ and $I(T)$ follow
the same $T$-dependence (see inset in Fig. 5), indicating that all the Cr$%
^{3+}$ ions that contribute to $\chi (T)$ also participate in $I(T)$.
However, for $T\lesssim $ $100$ K, $\chi (T)$ deviates from $I(T)$, and they
show maximums at $T\approx $ $40$ K and $T\approx $ $100$ K, respectively
(see inset in Fig. 5). The maximum in $\chi (T)$ is caused by AFM
correlations and indicates the onset of long range AFM ordering. Instead,
the maximum in $I(T)$ can be interpreted as transitions within thermally
populated exited states. The observation of EPR resonances in exited state
levels of nearest-neighbor Cr$^{3+}$ spin-coupled pairs diluted in the
spinel ZnGa$_{2}$O$_{4}$ has been reported by Henning et. al,. \cite{Henning}
These authors were able to determine, from the observed $I(T)$, the energy
separation between the first exited triplet state (S $=1$) and the ground
singlet state (S $=0$) to be $\left| J/k\right| \approx $ $32$ K. Also, from
the optical spectra of the Cr$^{3+}$ spin-coupled pairs in ZnGa$_{2}$O$_{4}$
a value of $\approx $ $32$ K was measured for $\left| J/k\right| $. \cite
{Gorkom} Within the same scenario and taking into account the thermal
population of all the exited states for the Cr$^{3+}$ spin-coupled pairs ($|%
{\bf S_{1}}+{\bf S_{2}}|$ - $|{\bf S_{1}}-{\bf S_{2}}|$; 3, 2, 1, and 0; $%
{\bf S_{1}}={\bf S_{2}}$ $=3/2$), the expected $T$-dependence of the total
EPR intensity, $I(T)$, in the three exited states levels at energies $J$ $%
,2J $, and $6J$ above the singlet ground state is given by:\cite{Henning}

\begin{equation}
I(T)\sim \left[ A\exp (-J/kT)+B\exp (-3J/kT)+C\exp (-6J/kT)\right] /Z
\end{equation}
where $Z=1+3exp(-J/kT)+5exp(-3J/kT)+7exp(-6J/kT)$ is the partition function,
the coefficients $A$, $B$, and $C$ are adjustable parameters proportional to
the transition probability within each excited multiplet. The solid line in
the inset of the Figure 4 shows the $T$-dependence given by Eq. 2 for $%
A=10(2)$, $B=1(0.5)$, $C=4500(200)$ and $J/k$ $=-45(2)$ K. The value found
for $\left| J/k\right| $ is larger than the one ($\approx $ $32$ K) found
for isolated Cr$^{3+}$ spin-coupled pairs in ZnGa$_{2}$O$_{4}$.\cite{Henning}
This difference is probably due to the different lattice parameters of ZnCr$%
_{2}$O$_{4}$ ($a$ $=8.327$ \AA ) in ZnGa$_{2}$O$_{4}$ ($a$ $=8.37$ \AA ). 
\cite{Boom} Nevertheless, the value is in good agreement with that extracted
from the Curie-Weiss temperature, $\Theta _{CW}$, (see above).

As we pointed out elsewhere \cite{Martinho}, one can also attribute the difference between the temperature dependence of the susceptibility and the EPR intensity to the presence of non--resonant low frequency modes that contribute spectral weight to the Kramers--Kronig integral for the static susceptibility,

\begin{equation}
\chi(T)=\frac{2}{\pi}\int_0^\infty d\omega \frac {\chi''(\omega)}{\omega},
\end{equation}
but do not participate in the EPR absorption. It is likely that such modes are seen in inelastic neutron scattering above $T_N$\cite{S.-H. Lee}. The fact that the susceptibility and the EPR intensity have a common temperature variation in the Cd--doped samples suggests that the non--resonant, low frequency modes, if present, are not making a significant contribution to the susceptibility integral.

Figure 7 shows the $C_{v}/T$ plots obtained from the data of Figure 6 for
each studied sample. The contribution from the magnetic component is
obtained from the difference with the data for the non magnetic reference
compound ZnGa$_{2}$O$_{4}$. The entropy, $S$, is obtained integrating these
differences and gives approximately the multiplicity of the involved levels, 
$\approx $ $2^{4}=16$. Within the same scenario of Cr$^{3+}$ spin-coupled
pairs, the Schottky anomaly for the spin-coupled pairs is given by: 
\begin{equation}
\frac{C_{v}}{T}=\frac{R}{T}\frac{\partial }{\partial T}\left[ T^{2}\frac{%
\partial \ln Z}{\partial T}\right]
\end{equation}
Figure 8 shows the fitting of the data to Eq. 3. The obtained value for $%
\left| J/k\right| $ $=35(2)$ K suggests the scheme of levels shown in Figure
8. The value found for $\left| J/k\right| $ is consistent with those values
obtained independently from the Curie-Weiss temperature, $\Theta _{CW}$, and
EPR intensity measurements in the HTPP (see above).

In the LTOP our EPR experiments show the appearance of small resonance modes
(see Figure 1 for $T\lesssim $ $17$ K). We believe that their sample size
and shape dependence and angular variation are, probably, more related to
demagnetizing effects rather than to the tetragonal crystal distortion
observed at $T\approx 12$ K.\cite{Hartmann} Besides, for $T\lesssim T_{N}$
the magnetic susceptibility increases at higher fields (see inset of Figure
3) and also, a small increase is oberved for $T\lesssim $ $5$ K (see Figure
3). These behaviors are similar to those observed in magnetization
measurements of polycrystalline samples of ZnCr$_{2}$O$_{4}$ and they have
been attributed to the presence of AFM domains in the LTOP.\cite{D. Fiorani}
Thus, we associate our resonance modes with AFM domains that might be present
in these materials as a consequence of their highly frustrated 3D long range
AFM magnetic structure.

\section{\bf CONCLUSIONS}

In conclusion, our EPR and $\chi (T)$ results in the $normal$ spinel ZnCr$%
_{2}$O$_{4}$ show, between $12$ K and $100$ K, a transition from a long to a
short range regime of AFM correlations (LTOP-HTPP). From the $T$-dependence
of the EPR intensity in the HTPP an exchange parameter of $J/k\approx $ $-45$
K between the Cr$^{3+}$ (S $=3/2$) spin-coupled pairs was extracted. This
value is close to the one obtained independently from the Curie-Weiss
temperature, $\Theta _{CW}$, and from the Schottky anomaly observed in the
specific heat. Thus, the magnetic properties of these strongly frustrated
systems in the HTPP can be described within a scenario involving just
spin-coupling pairs of Cr$^{3+}$ (S $=3/2$). The sharp drop in $\chi (T)$ at 
$T\approx $ $12$ K, the peak in $C_{v}(T)$ also at $T\approx $ $12$ K, and
the ordering temperature extracted from the broadening of the EPR linewidth (%
$T_{N}\approx $ $12$ K) confirmed the AFM ordering at $T\approx $ $12$ K in
ZnCr$_{2}$O$_{4}$. The resonance modes observed in the LTOP and the field
dependent susceptibility, $\chi (T,H)$, indicates the presence of AFM
domains in this material. Finally, we found that the disorder caused by the
Cd impurities in ZnCr$_{2}$O$_{4}$ drives the system from an AFM to a SG
type of highly frustrated magnetic ordering.

Although a model based on isolated pairs can account for many of the magnetic properties of ZnCr$_2$O$_4$, it does not include the interaction of the pairs with the surrounding ions. A widely used pair model that does include the effects of the interaction with neighboring spins is the constant coupling approximation \cite{Kasteleijn}. However, this model does not exploit the unique tetrahedral character of the Cr sublattice in ZnCr$_2$O$_4$. In particular, it predicts the same susceptibility for ZnCr$_2$O$_4$ as found in an unfrustrated simple cubic antiferromagnet, which has the same number of nearest neighbor interactions (6) and exhibits a conventional AFM transition. This contradiction has led two of us (AJGA and DLH) to develop a quantum tetrahedral mean field model \cite{Garcia}. In this model, the energy levels of a four--spin tetrahedral cluster are calculated exactly and the interaction with the neighboring ions is treated in the mean field approximation. Good agreement with the susceptibility and the magnetic specific heat are obtained with nearest--neighbor interaction, $J_1=38.6$ K, and next--neighbor interaction, $J_2=1.4$ K\cite{Martinho,Garcia}.

\section{Acknowledgments}

This work was supported by FAPESP\ Grants No 95/4721-4, 96/4625-8,
97/03065-1, and 97/11563-1 S\~{a}o Paulo-SP-Brazil, and NSF-DMR No 9705155,
and NSF-INT No 9602928.  AJGA wants to thank the Spanish MEC for financial support under the Subprograma General de Formaci\'{o}n de Personal Investigador en el Extranjero.

\pagebreak
\pagestyle{empty}

\begin{figure}[p]
\resizebox*{!}{0.9\textheight}{\includegraphics{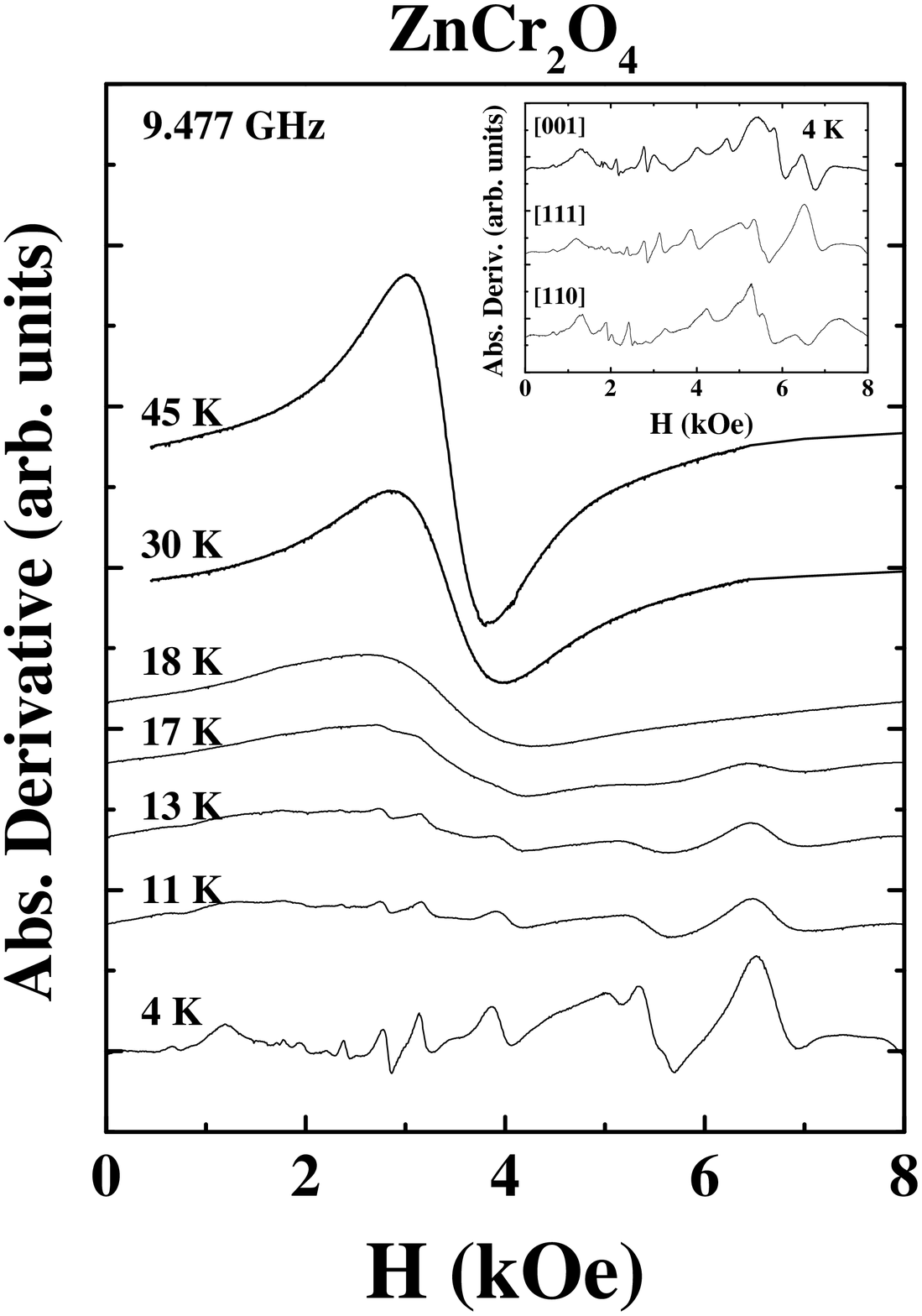}}
\caption{$T$-evolution ($4$ K $\leqslant $ $T$ $\leqslant 45$ K) of the EPR
spectra for a ZnCr$_{2}$O$_{4}$ single crystal. The inset shows the spectra
at $T=4$ K for different field orientations in the (110) plane.}
\end{figure}
\begin{landscape}
\begin{figure}
\scalebox{0.8}{\includegraphics{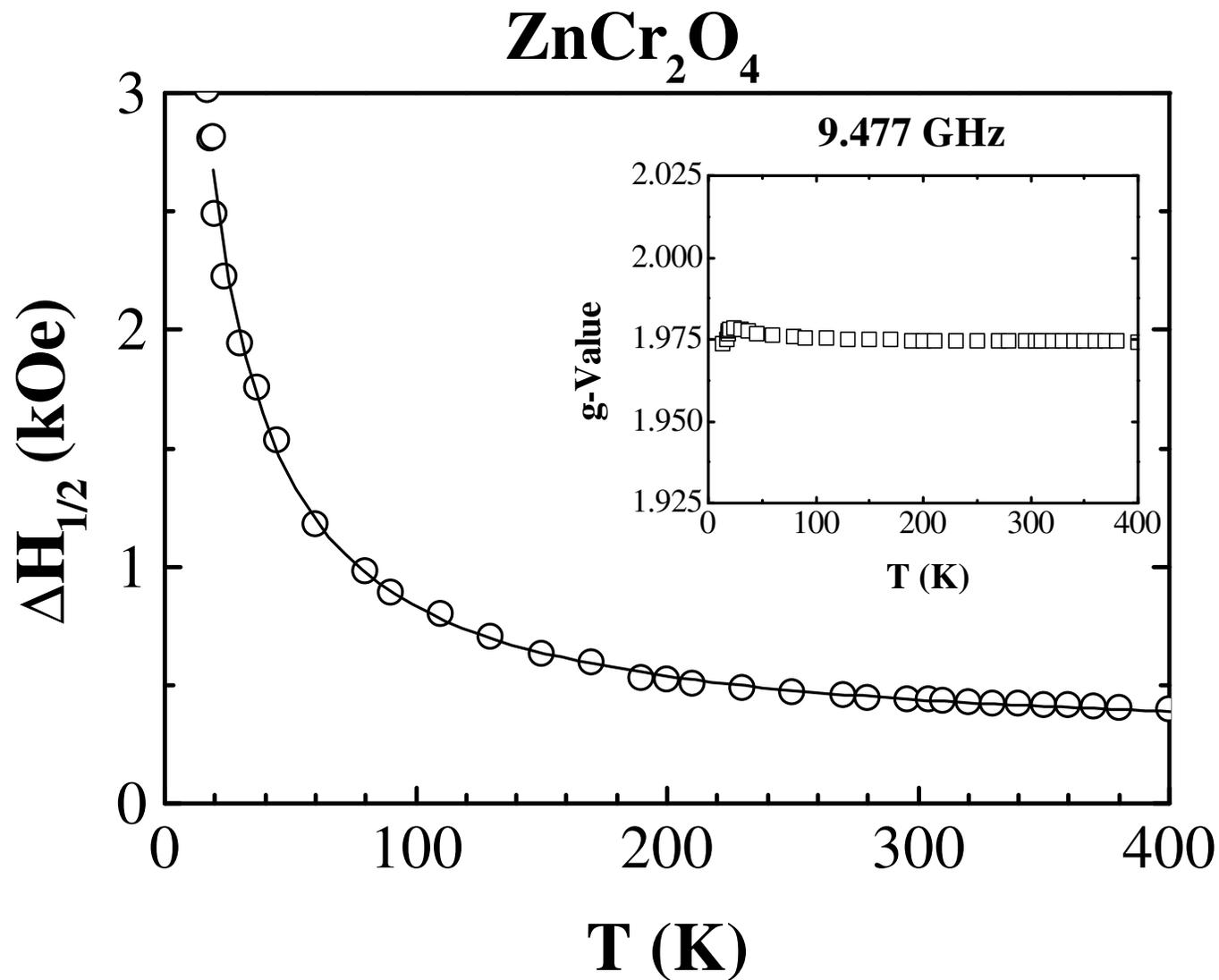}}
\caption{$T$-dependence ($18$ K $\leqslant $ $T$ $\leqslant $ $400$ K) of
the EPR linewidth and $g$-value for the crystal of Fig. 1. The solid line
shows the best fit of the linewidth to Eq. 1 (see text). }
\end{figure}

\begin{figure}[p]
\scalebox{0.85}{\includegraphics{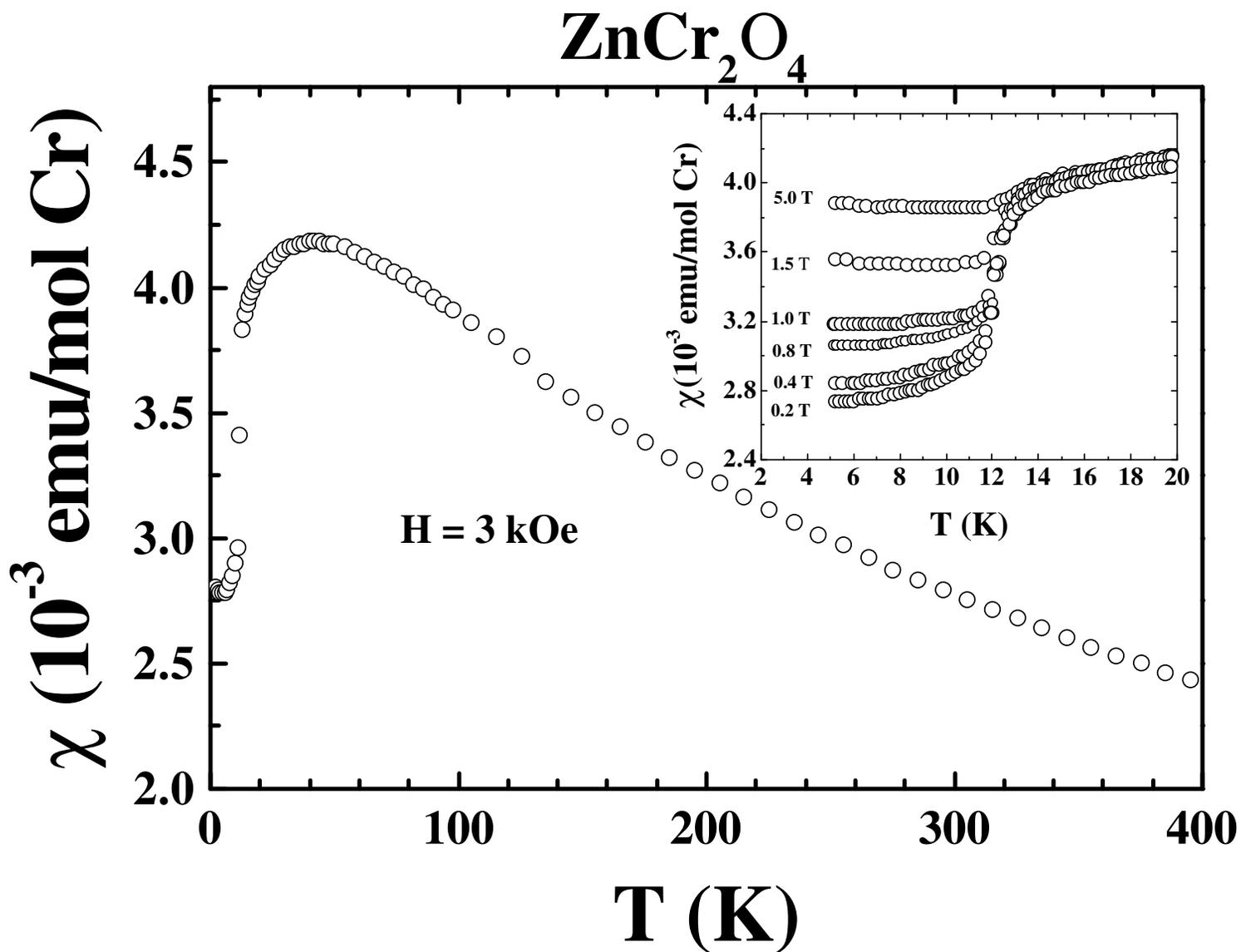}}
\caption{$T$-dependence ($2$ K $\leqslant $ $T$ $\leqslant $ $400$ K) of the
magnetic susceptibility, $\protect\chi (T)$, at $H$ = $3$ kOe (FC, ZFC). The
inset shows the data for $2$ K $\leqslant $ $T$ $\leqslant $ 20 K and $0.2$
T $\leqslant $ $H$ $\leqslant 5$ T.}
\end{figure}

\begin{figure}[p]
\scalebox{0.85}{\includegraphics{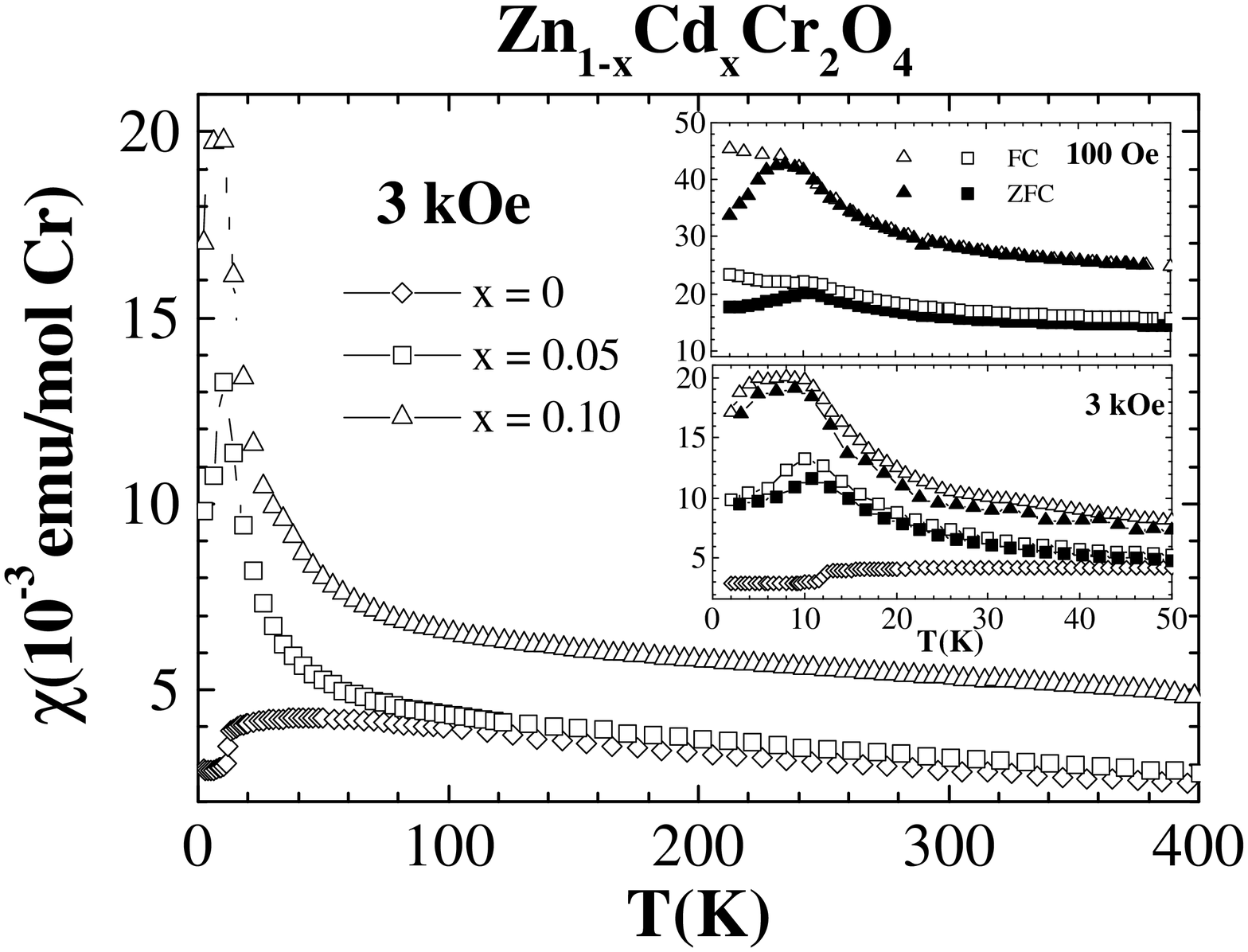}} 
\caption{$T$-dependence ($2$ K $\leqslant $ $T$ $\leqslant $ $400$ K) of the
FC magnetic susceptibility, $\protect\chi (T)$, at $3$ kOe for Zn$_{1-x}$Cd$%
_{x}$Cr$_{2}$O$_{4}$. The insets shows FC and ZFC data for $2$ K $\leqslant $
$T$ $\leqslant $ $50$ K at $H=100$ Oe and $3$ kOe.}
\end{figure}

\begin{figure}[p]
\scalebox{0.8}{\includegraphics{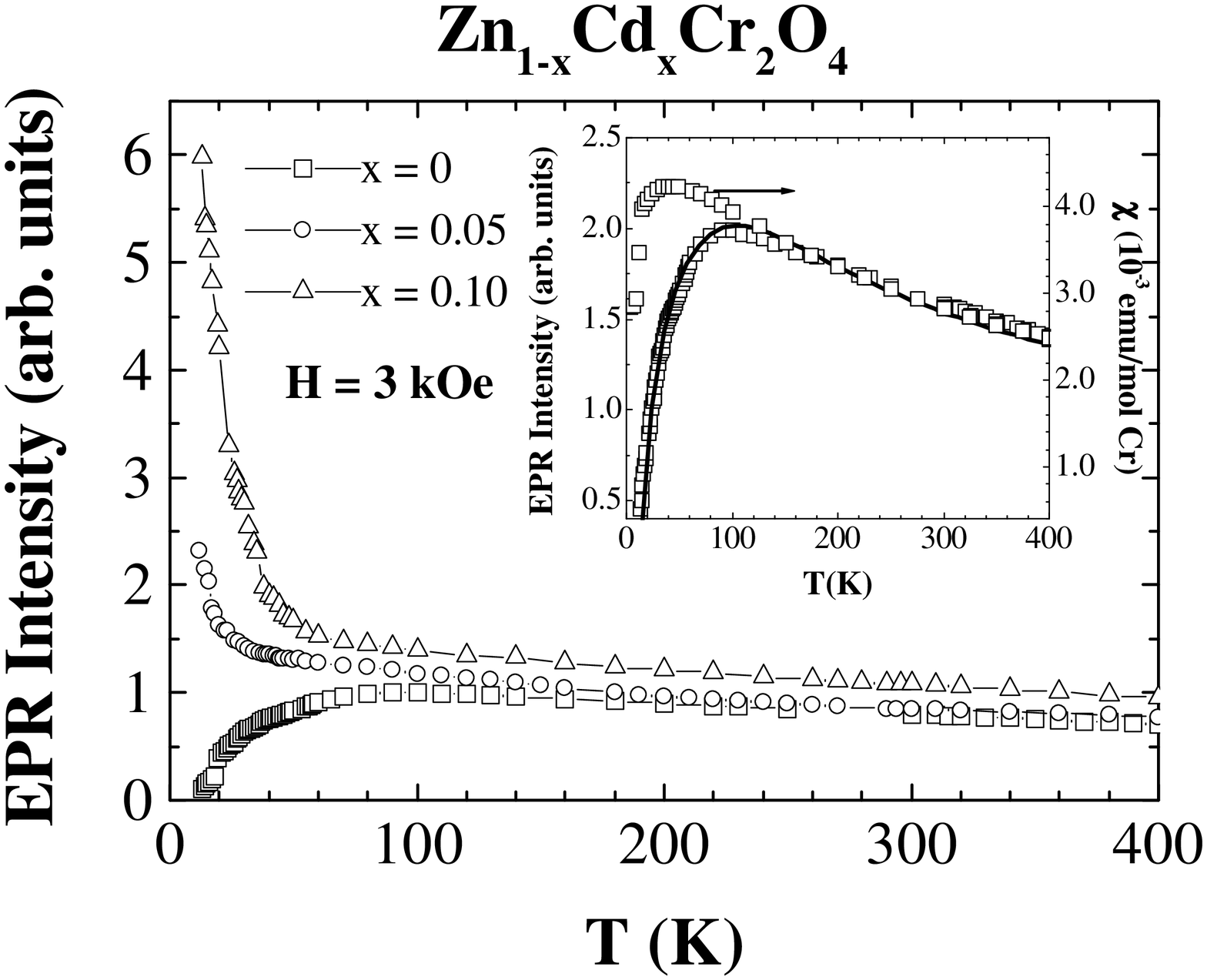}} 
\caption{$T$-dependence ($15$ K $\leqslant $ $T$ $\leqslant $ $400$ K) of
the EPR intensity, $I(T)$ for Zn$_{1-x}$Cd$_{x}$Cr$_{2}$O$_{4}$. The inset
compares $I(T)$ and $\protect\chi (T)$ for ZnCr$_{2}$O$_{4}$ and the solid
line is the best\ fit of $I(T)$ to Eq. 2 (see text).}
\end{figure}

\begin{figure}[p]
\scalebox{0.85}{\includegraphics{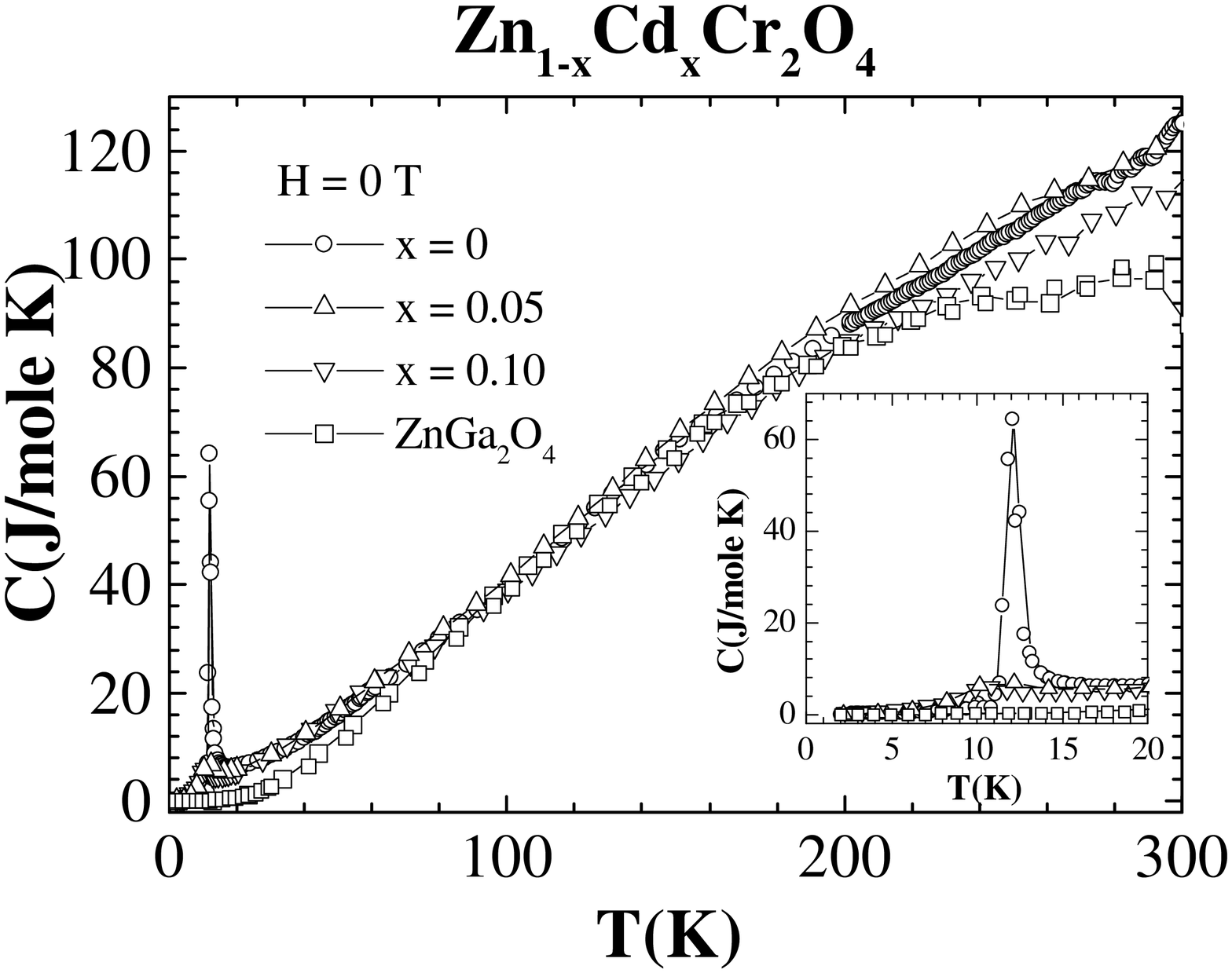}} 
\caption{$T$-dependence ($2$ K $\leqslant $ $T$ $\leqslant $ $300$ K) of the
ZF specific heat, $C_{v}(T)$, for Zn$_{1-x}$Cd$_{x}$Cr$_{2}$O$_{4}$ and ZnGa$%
_{2}$O$_{4}$. The inset shows $C_{v}(T)$ for $2$ K $\leqslant $ $T$ $%
\leqslant $ $300$ K.}
\end{figure}
\end{landscape}

\begin{figure}[p]
\resizebox*{!}{0.9\textheight}{\includegraphics{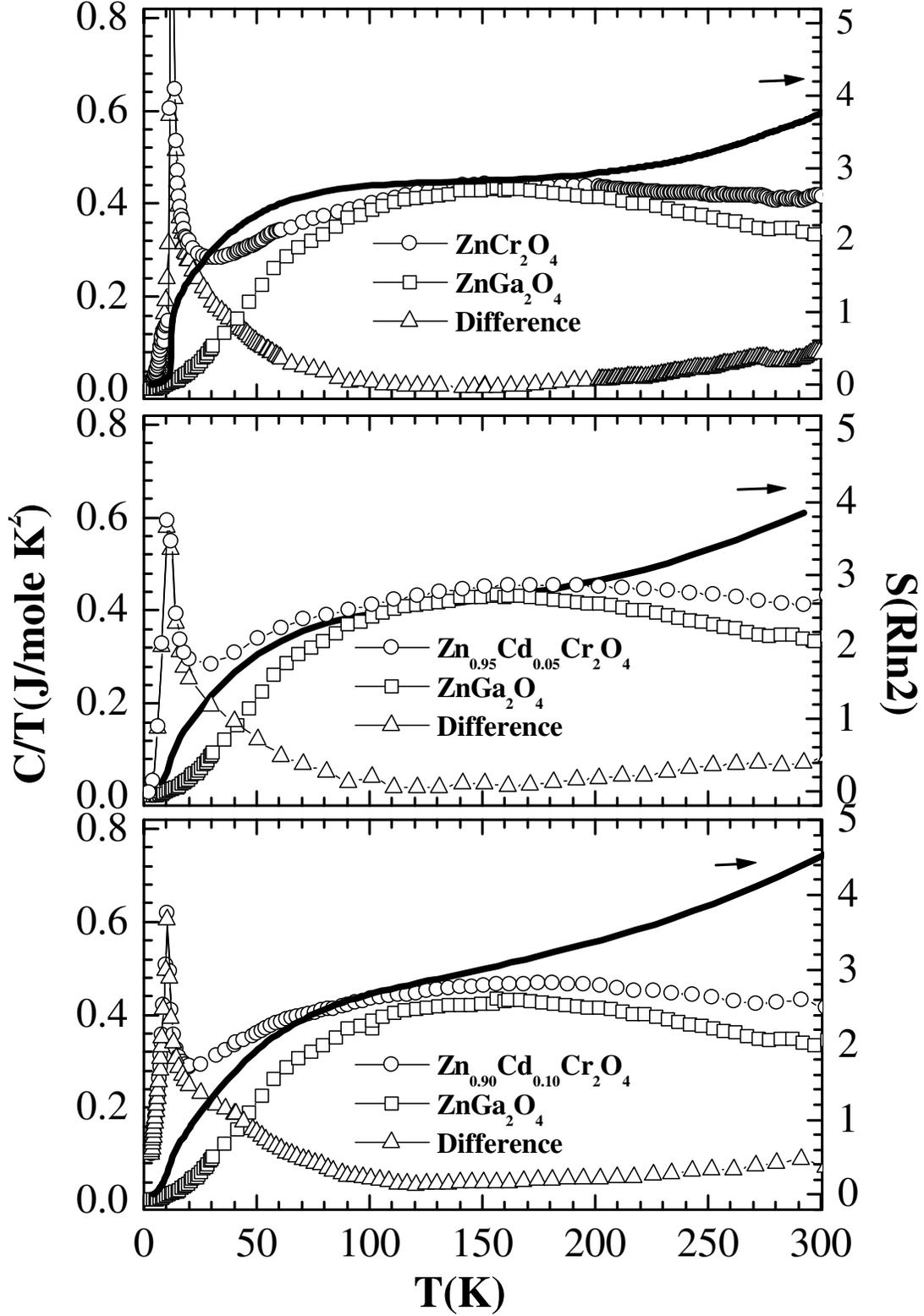}}
\caption{$T$-dependence ($2$ K $\leqslant $ $T$ $\leqslant $ $300$ K) of the
ZF $C_{v}/T$ \ for Zn$_{1-x}$Cd$_{x}$Cr$_{2}$O$_{4}$ and ZnGa$_{2}$O$_{4}$.
The difference with the reference compound is also shown. The solid line
gives the entropy, $S$, associated to the system's level multiplicity. }
\end{figure}

\begin{landscape}
\begin{figure}[p]
\scalebox{0.8}{\includegraphics{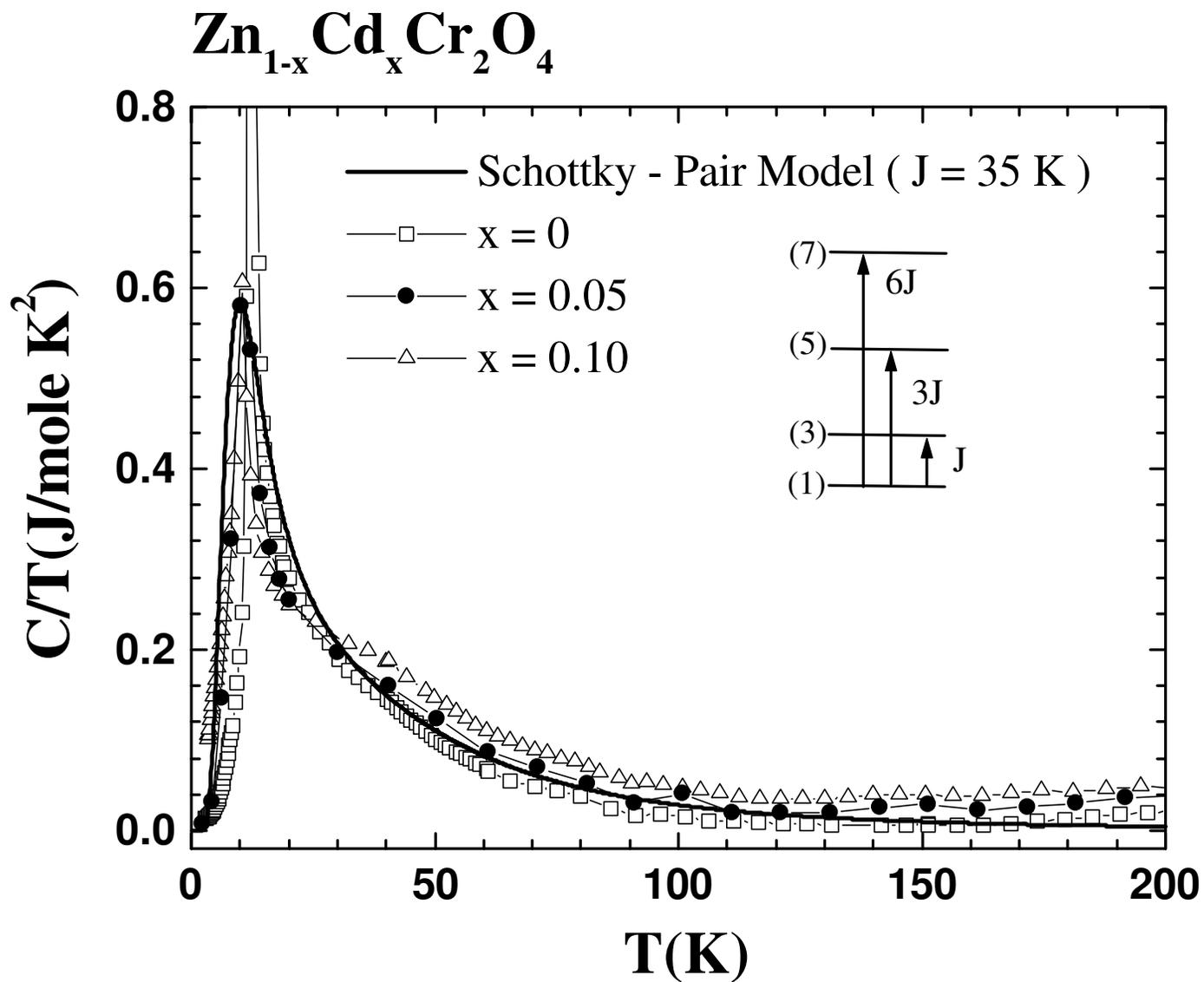}} 
\caption{$T$-dependence ($2$ K $\leqslant $ $T$ $\leqslant $ $300$ K) of the
magnetic contribution to \ $C_{v}/T$ . The dashed line is the Schottky
anomaly calculated using Eq. 4 for a pair model of coupled Cr$^{3+}$ spins
(see text). }
\end{figure}
\end{landscape}

\narrowtext
\begin{table}[p]
\caption{Curie-Weiss parameters for Zn$_{1-x}$Cd$_{x}$Cr$_{2}$O$_{4}$}
\label{table1}
\begin{tabular}{cccc}
Zn$_{1-x}$Cd$_{x}$Cr$_{2}$O$_{4}$ & $C$ & $\theta _{CW}$ & $\mu _{eff}$ \\ 
& $($emu/mole Cr K$)$ & $($K$)$ & $(\mu _{B})$ \\ 
\tableline$x=0.0$ & 1.95(2) & -390(20) & 3.95(10) \\ 
\tableline$x=0.05$ & 2.94(5) & -500(20) & 4.85(20) \\ 
\tableline$x=0.10$ & 2.57(5) & -483(20) & 4.53(20)
\end{tabular}
\end{table}


\begin{references}
\bibitem{Anderson}  P.W. Anderson, Phys. Rev. {\bf 79}, 350 (1950); ibid, 
{\bf 79}, 705 (1950); ibid, {\bf 102}, 1008 (1956).

\bibitem{De Seze}  L. De Seze, J. Phys. C, Solid State Phys. {\bf 10}, L353
(1977); J. Villain, Z. Phys. B {\bf 33}, 31 (1979).

\bibitem{Fiorani}  D. Fiorani, S. Viticoli, J.L. Dormann, J.L. Tholence, J.
Hammann, A.P. Murani, and J. Soubeyroux, Solid State Phys. {\bf 16}, 3175
(1983); D. Fiorani, J. Phys. C {\bf 17}, 4837 (1984).

\bibitem{Barth}  T.F.W. Barth and E. Posnjak, S. Krist. {\bf 82}, 325 (1932).

\bibitem{Moessner}  R. Moessner and J.T. Chalker, Phys. Rev. Lett. {\bf 80},
2929 (1998); R. Moessner, Phys. Rev. B {\bf 57}, R5587 (1998).

\bibitem{Canals}  B. Canals and C. Lacroix, Phys. Rev. Lett. {\bf 80}, 2933
(1998).

\bibitem{Ramirez}  A.P. Ramirez, G.P. Espinosa, and A.S. Cooper, Phys. Rev.
B {\bf 45}, 2505 (1992).

\bibitem{Plumier}  R. Plumier, M. Lecomte, and M. Sougi, J. de Physique
Letters {\bf 38}, L149 (1977).

\bibitem{Hartmann}  F. Hartmann-Boutron, A. Gerard, P. Imbert, R.
Kleibergerard, and F. Varret, Comptes Rendus de L Academie des Sciences
Paris, Serie II Fascicule B {\bf 268}, 906 (1969); S.-H. Lee, C. Broholm,
T.h. Kim, W. Rotcliff, S.-W. Cheong, and Q. Huang, unpublished;

\bibitem{S.-H. Lee}  S.-H. Lee, C. Broholm, T.H. Kim, W. Ratcliff, and S.-W.
Cheong, Phys. Rev. Lett. {\bf 84}, 3718 (2000).

\bibitem{Cheong}  Sample preparation

\bibitem{Baran}  M. Baran, S. Piechota, and A. Pajaczkowska, Acta Phys.
Polon. Part A {\bf 59}, 47 (1981).

\bibitem{O'Reilly}  D.E. O'Reilly and D.S. MacIver, J. Phys. Chem. {\bf 66},
276 (1962).

\bibitem{Forni}  L. Forni and C. Oliva, J. Chem. Soc. Faraday Trans. 1 {\bf %
84}, 2477 (1988).

\bibitem{Boom}  H. van den Boom, J.C.M. Henning, and J.P.M. Damen, Solid
State Commun. {\bf 8}, 717 (1970).

\bibitem{Huber}  D.L. Huber, Phys. Rev. B {\bf 6}, 3180 (1972); M.S. Seehra,
Phys. Rev. B {\bf 6}, 3186 (1972); K. Kawasaki, Prog. Theor. Phys. {\bf 39},
285 (1968).

\bibitem{D. Fiorani}  D. Fiorani and S. Viticoli, J. Magnetism and Ma gnetic
Materials, {\bf 49}, 83 (1985); L. N\'{e}el, Proceed. Int. Conf. Theory.
Phys. (Kyoto) (1954).

\bibitem{Henning}  J.C.M. Henning, J.H. den Boef, and G.G.P. van Gorkom,
Phys. Rev. B {\bf 7}, 1825 (1973).

\bibitem{Gorkom}  G.G.P. Gorkom, J.C.M. Henning, and R.P. van Stapele, Phys.
Rev. B {\bf 8}, 955 (1973).

\bibitem{Martinho} H. Martinho, N.O. Moreno, J.A. Sanjuro, C. Rettori, A.J. Garc\'{i}a--Adeva, D.L. Huber, S.B. Oseroff, W. Ratcliff, II, S.-W. Cheong, P.G. Pagliuso, J.L. Sarrao, and G.B. Martins, paper DH-08, 8th Joint MMM - Intermag Conference, San Antonio, Jan. 2001.

\bibitem{Kasteleijn} P.W. Kasteleijn and J.H. Van Kranendonk, Physica \textbf{22}, 317 (1956).

\bibitem{Garcia} A.J. Garc\'{i}a--Adeva and D.L. Huber, Phys. Rev. Lett. \textbf{85}, 4598 (2000). 

\end{references}
\end{document}